\documentclass{PoS}
\usepackage{cite}
\usepackage{amsmath}
\usepackage{bm}
\usepackage[pdftex]{graphicx}
\usepackage{subfigure}
\title{Universal scaling of non-equilibrium critical fluctuations from Langevin dynamics of model A}

\ShortTitle{Universal scaling of Langevin dynamics}

\author{Shanjin Wu\\
        Department of Physics and State Key Laboratory of Nuclear Physics and
Technology, Peking University, Beijing 100871, China\\
        E-mail: \email{shanjinwu2014@pku.edu.cn}}

\author{Huichao Song\\
        Department of Physics and State Key Laboratory of Nuclear Physics and
Technology, Peking University, Beijing 100871, China\\
        Collaborative Innovation Center of Quantum Matter, Beijing 100871, China\\
        Center for High Energy Physics, Peking University, Beijing 100871, China\\
        E-mail: \email{huichaosong@pku.edu.cn}}

\abstract{Within the framework of the Kibble-Zurek Mechanism, we investigate the universal behavior of the non-equilibrium critical fluctuations, using the Langevin dynamics of model A. With properly located typical time, length and angle scales, $\tau_{\mbox{\tiny KZ}}$, $l_{\mbox{\tiny KZ}}$, and $\theta_{\mbox{\tiny KZ}}$, the constructed functions $\bar{f}_n((\tau-\tau_c)/\tau_{\mbox{\tiny KZ}},\theta_{\mbox{\tiny KZ}})$ (n=1...4) for the cumulants of the sigma field show universal behavior near the critical point, which are independent from some non-universal factors, such as the relaxation time or the evolution trajectory.}

\FullConference{Critical Point and Onset of Deconfinement - CPOD2017\\
		7-11 August, 2017\\
		The Wang Center, Stony Brook University, Stony Brook, NY}

\begin{document}

\section{Introduction}

Recently, the RHIC Beam Energy Scan (BES)  program  has measured the higher order cumulants of net charges and net protons in Au+Au collisions at different collision energies ranging from 7.7 to 200 A GeV~\cite{Aggarwal,Adamczyk,Luo,Adamczyk:2014fia,Thader:2016gpa}. The kurtosis of net protons shows a large deviation from the poisson baseline and presents obvious non-monotonic behavior from lower to higher collision energies, which indicates the potential of locating the QCD critical point.

Although equilibrium critical fluctuation could explain the acceptance dependence for the cumulants of the net protons \cite{Stephanov:2008qz,Jiang:2015hri,Ling:2015yau}, it fails to qualitatively describe the skewness data due to the intrinsic positive contributions~\cite{Jiang:2015hri}.  On the other hand,  it is
also important to address the non-equilibrium effects near the critical point due to the dramatically increased relaxation time of the slow mode. In Ref.~\cite{Mukherjee:2015swa,Jiang:2017mji}, the non-equilibrium critical fluctuation of the sigma field have been investigated using Fork Planck equations and Langevin dynamics within the framework of model A, which showed that the critical slowing down effects could change the signs of the higher order cumulants compared with equilibrium ones.

These predicted cumulants of the non-equilibrium critical fluctuation are influenced by free inputs and free parameters in the model calculations, such as the evolution trajectory of the heat bath, the relaxation time and damping coefficient, initial conditions, etc. Recently, it was realized that, the Kibble-Zurek Mechanism (KZM) could lead to the emergence of the universal scaling for a dynamical evolving system that undergoes a continuous phase transition with the critical slowing down. The KZM was first pointed out by Kibble~\cite{Kibble:1976} in cosmology and then extended to the condensed matter physics by Zurek~\cite{Zurek:1985}. After then, the KZM has been applied to various non-equilibrium systems with both classical~\cite{Chandran:2012} and quantum phase transitions~\cite{Polkovnikov:2005,Zurek:2005,Dziarmaga:2005}. In
relativistic heavy ion collisions, the KZM has been applied to the non-equilibrium dynamics of the Fokker-Planck equation and extracted several universal scaling functions for the cumulants of the zero mode sigma field~\cite{Mukherjee:2016kyu},
which indicates the possibility of constructing parameter-independent observables for the search of the QCD critical point.
In this proceeding, we will investigate the universal behavior of the critical fluctuations with Langevin dynamics of model A, which keeps the spatial information of the evolving sigma fielld.

\section{Langevin dynamics near QCD critical point}

As a system evolves near the critical point,  its non-equilibrium dynamics can be described by a dynamical model associated with certain dynamical universality class. It is generally believed that the evolving hot QCD system belongs to model H~\cite{Son:2004iv} according to the classification of~\cite{Rev1977}. However, the related numerical simulations are  complicated, which have not been implemented so far. For simplicity, we focus on the dynamics of model A, which only evolves the non-conserved order parameter field. For the Langevin dynamics of model A,  the evolution equation can be written as:
\begin{align}\label{Langevin3+1}
  \frac{\partial \sigma(\tau,\bm{x})}{\partial \tau} = - \frac{1}{m^2_\sigma \tau_{\mbox{\tiny eff}}} \frac{\delta U [\sigma(\bm{x})]}{\delta \sigma(\bm{x})} + \zeta(\bm{x},\tau),
\end{align}
where  $\tau_{\mbox{\tiny eff}}$ is the relaxation time,  $m_\sigma$ is the mass of the sigma field and $\zeta(\bm{x},\tau)$ is the white noise that satisfies the fluctuation-dissipation theorem.  The effective potential $U(\sigma)$ can be expanded in the powers of $\sigma(\bm{x})$:
\begin{align}
\begin{aligned}
  U[\sigma]= \int d^3x&\left\{\frac{1}{2}[\nabla\sigma(\bm{x})]^2+\frac{1}{2}m^2_\sigma[\sigma(\bm{x})-\sigma_0]^2+\frac{\lambda_3}{3}[\sigma(\bm{x})-\sigma_0]^3+\frac{\lambda_4}{4}[\sigma(\bm{x})-\sigma_0]^4\right\}.
\end{aligned}
\end{align}
where $\sigma_0$ is the equilibrium mean value of $\sigma(x)$,  $\lambda_3$ and $\lambda_4$
are the coupling coefficients, and the mass $m_{\sigma}$ and the equilibrium correlation length $\xi_{eq}$ satisfies $\xi_{eq}=1/m_{\sigma}$.  These parameters can be obtained from identifying the equilibrium cumulants of this potential with the ones from the 3d Ising model, together with a linear map between the QCD variable $(T, \mu)$ and the Ising model variable ($r,h$): $(T-T_c)/\Delta T=h/\Delta h,\,(\mu-\mu_c)/\Delta \mu =-r/\Delta r$. Note that, in Ising model,  $r$ and $h$ can be expressed with  variables $R$ and $\theta$: $ r(R,\theta)= R(1-\theta),\,h(R,\theta) = R^{5/3}(3\theta-2\theta^3)$, from which the changing rate of $\theta$ is used to characterize the changing of the thermodynamic potential in the following calculations. In this proceeding, we set other related parameters as: $\Delta T=T_c/8$, $\Delta\mu=0.1\mbox{GeV}$, $\Delta r=(5/3)^{3/4}$, $\Delta h=1$, $T_c=0.16\mbox{GeV}$ and $\mu_c=0.395$GeV. For the detailed explanation, please refer to Ref.~\cite{Wu,Mukherjee:2015swa}.

\begin{figure}[tbp]
  \centering
  \subfigure{
  \begin{minipage}{4.6cm}
  \centering
   \includegraphics[width=5.2cm]{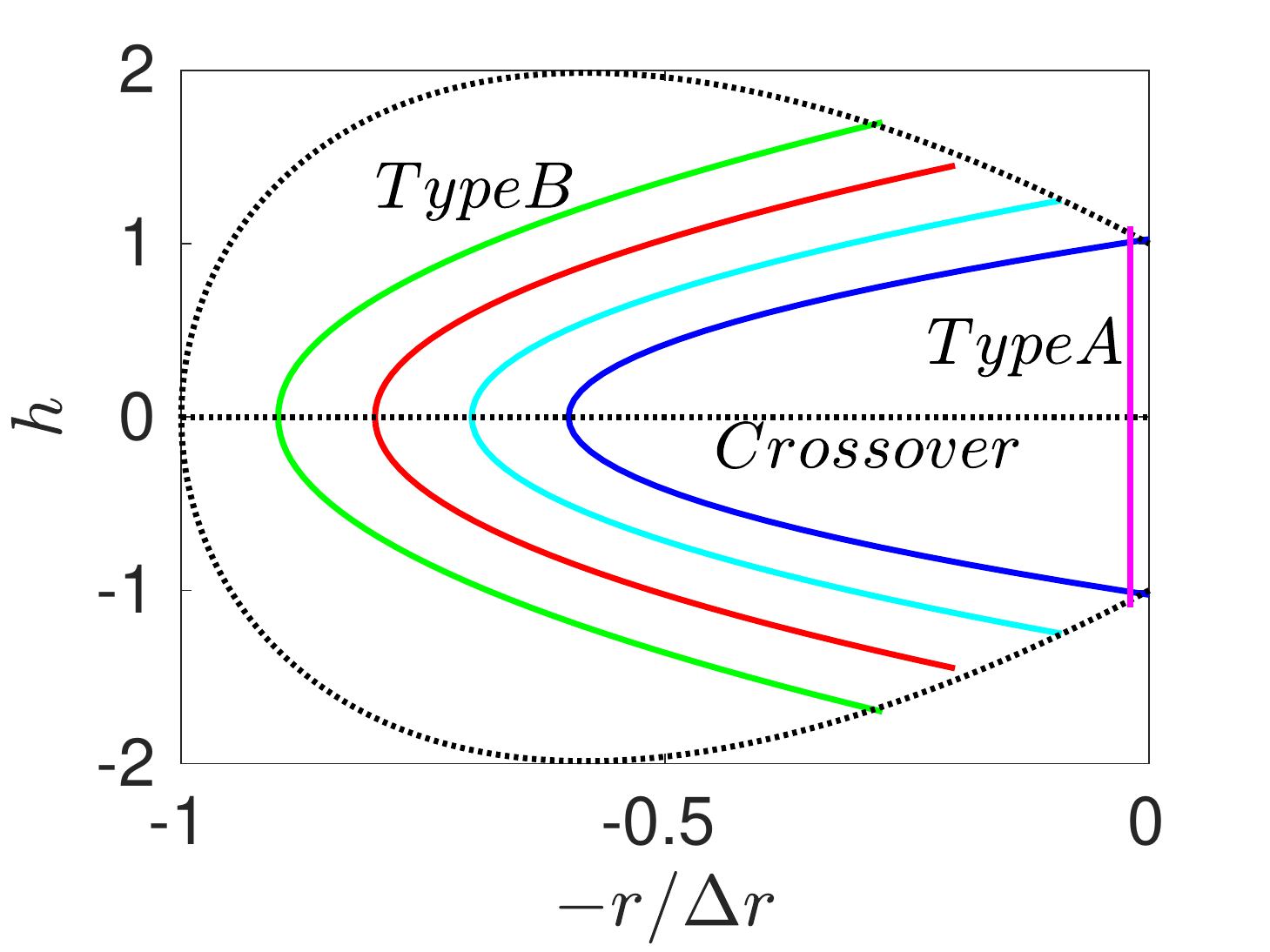}
  \end{minipage}
  }
\subfigure{
  \begin{minipage}{4.6cm}
  \centering
   \includegraphics[width=5.2cm]{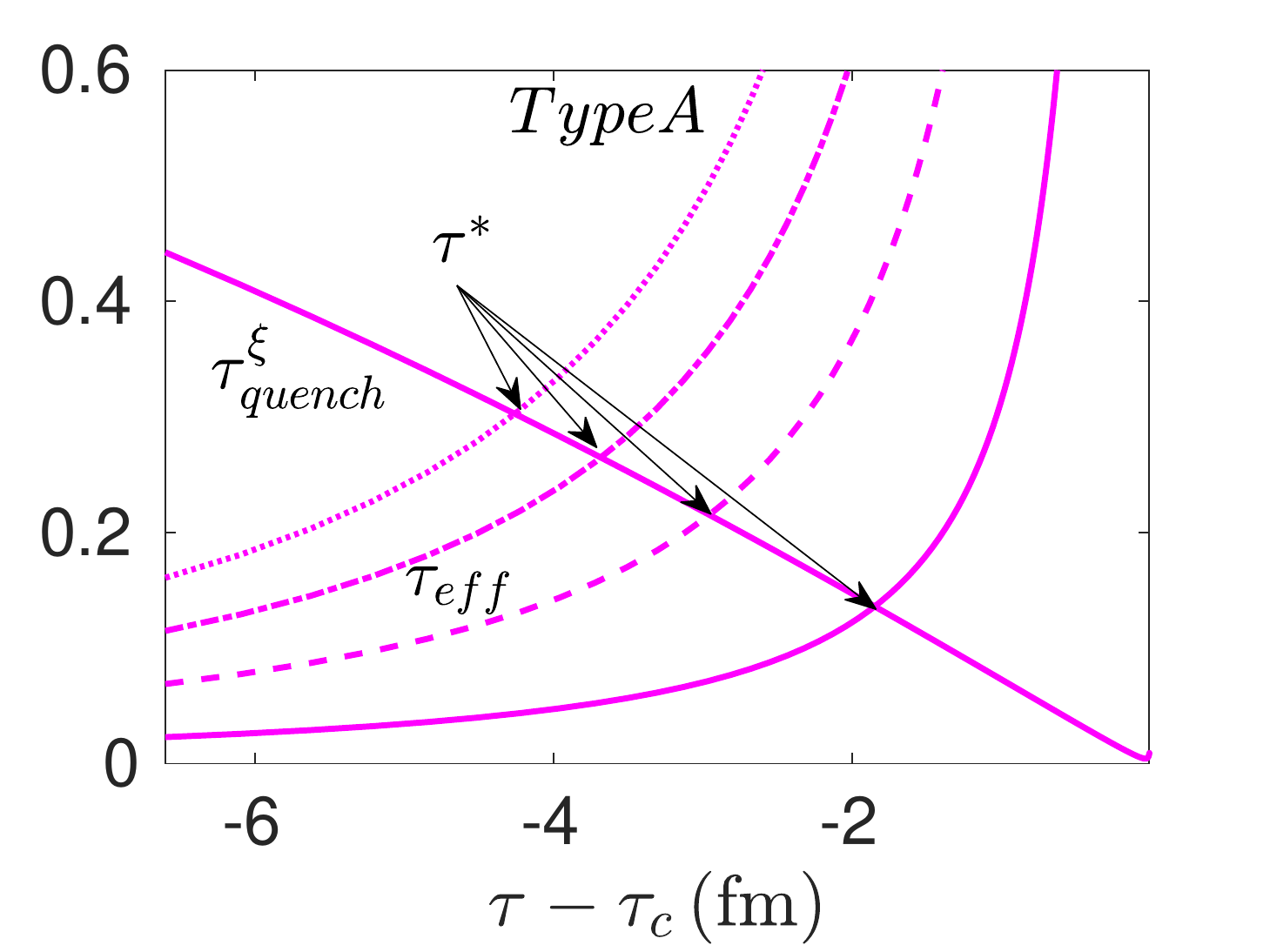}
  \end{minipage}
  }
\subfigure{
  \begin{minipage}{4.6cm}
  \centering
   \includegraphics[width=5.2cm]{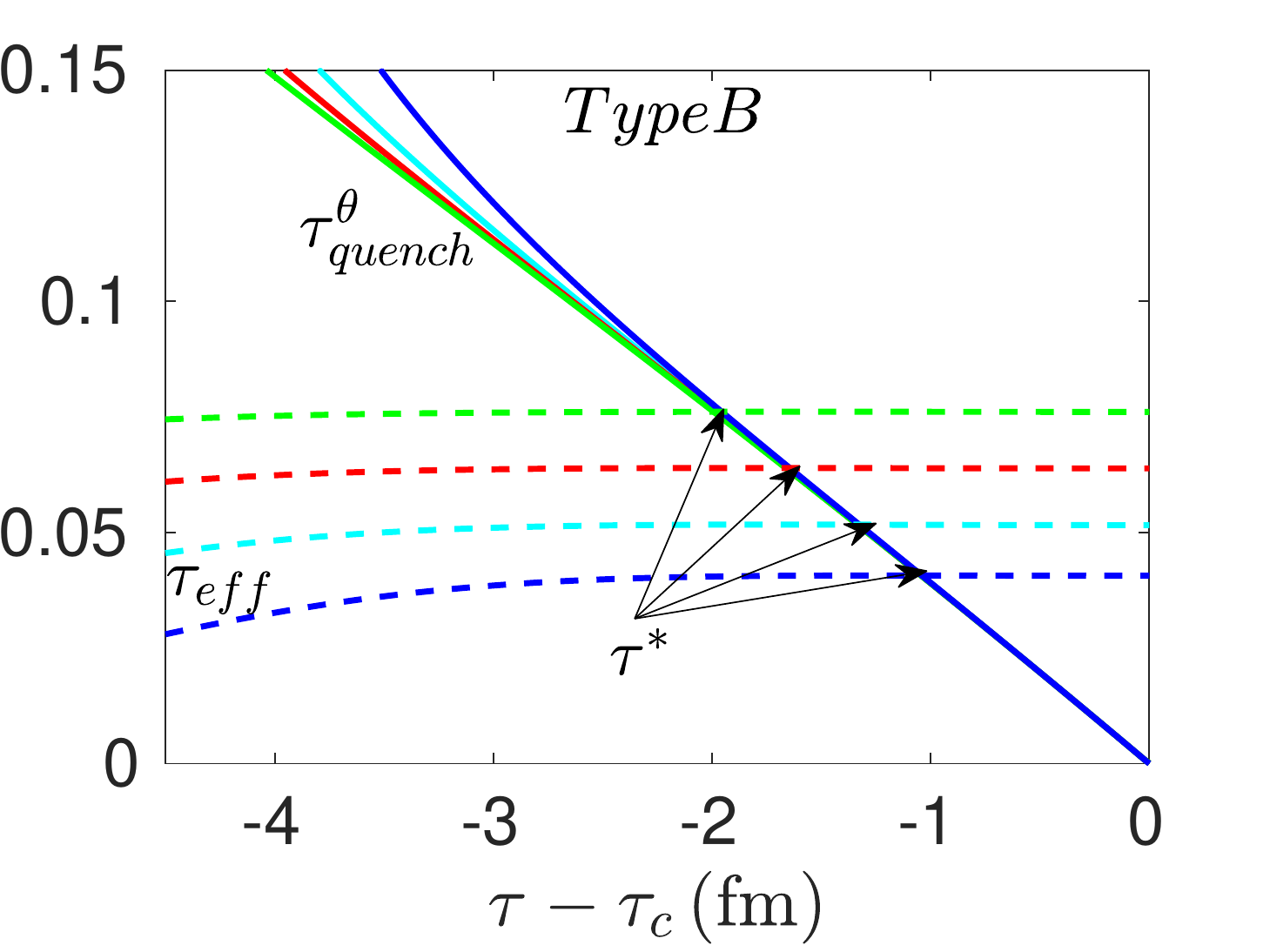}
  \end{minipage}
  }
  \caption{(color online) Left panel: illustrations of two types of trajectories on the phase diagram in $r-h$ plane.
  The magenta line is the trajectory of type A with a fixed chemical potential. Other colored curves are trajectories of type B  with approximately equal correlation length  near the crossover line.  The black dashed curve donates the boundary of critical regime with $\xi_{eq}=1$fm. Middle and right panels: illustration of locating $\tau^*$
  for type A and type B trajectories. The solid and dashed lines are quench times $\tau_{\mbox{\tiny quench}}$ and relaxation times $\tau_{\mbox{\tiny eff}}$, respectively. The points where $\tau_{\mbox{\tiny eff}}=\tau_{\mbox{\tiny quench}}$ gives the proper time $\tau^*$ and the corresponding $\tau_{KZ}$. }
  \label{fig2}
\end{figure}

To numerically solve Eq.~(\ref{Langevin3+1}), one needs to input the temperature $T(\bm{x})$ and chemical potential $\mu(\bm{x})$ profiles for the evolving potential $U[\sigma]$ with the $T-\mu$ dependent parameters $m_\sigma$,  $\lambda_3$ and $\lambda_4$. Here, we assume that $T(\bm{x})$ and $\mu(\bm{x})$ are provided by an external heat bath that evolves along two types of trajectories that satisfies: $r=r_c-a_h h^2$, where $r_c$ and $a_h$ are free parameters.  For the trajectory of type A, we set $a_h=0$ and $r_c=0.02\Delta r$. This corresponds to a trajectory with a fixed chemical potential, where the change of the thermodynamical potential is mainly controlled by the change of the correlation length $\xi_{eq}$.  The trajectories of type B are constructed with approximately equal correlation length  $\xi_{eq}$ near the crossover line, which ensures that the changing rate of $\theta$ is the dominant factor for the change of the thermodynamical potential. Here, we first set $r_c=0.6,0.7,0.8$ and $0.9$ and then tune $a_h$ in a way that leads to approximate equal $\xi_{eq}$ near the cross over phase transition.  The left panel of Fig.~\ref{fig2} shows the critical regime of the phase diagram in $r-h$ plane with two types of trajectories A and B, where the boundary of critical regime is defined by the correlation length $\xi_{eq}=1$ fm.

For both types of trajectories A and B, we assume that the temperature of the heat bath drops down in a Hubble-like way: $T=T_I(\tau/\tau_I)^{-0.45}$,  where $\tau_I$ and $T_I$ are the initial time and temperature, respectively. The effective relaxation time $\tau_{\mbox{\tiny eff}}$ is another necessary input of Eq.~(\ref{Langevin3+1}).  Here we set it as $\tau_{\mbox{\tiny eff}}= \tau_{\mbox{\tiny rel}}(\xi_{\mbox{\tiny eq}}/\xi_{\mbox{\tiny min}})^3$~\footnote{For the dynamical critical exponent, we use the one from model H with $z=3$.}, where $\xi_{\mbox{\tiny min}}$ and $\tau_{\mbox{\tiny rel}}$  represent the correlation length and the relaxation time at the boundary of the critical regime, and $\tau_{\mbox{\tiny rel}}$ is the free parameter in our calculations.  With the probability function: $ P\left[ \sigma\left( \bm{x}\right) \right] \sim \exp \left( -U \left( \sigma\right) /T\right) $, the initial profiles of the sigma field can be constructed, which are then evolved according to Eq.~(\ref{Langevin3+1}) event by event. With the obtained $\sigma\left( \bm{x}\right)$ profiles, we  calculate the time dependent cumulants: $C_1\equiv\langle\sigma\rangle,C_2\equiv\langle(\delta\sigma)^2\rangle,C_3\equiv\langle(\delta\sigma)^3\rangle,
C_4\equiv\langle(\delta\sigma)^4\rangle-3\langle(\delta\sigma)^2\rangle$, where the variance is defined as $\delta\sigma\equiv\sigma-\langle \sigma\rangle$, $\sigma$ denotes the spatial average of $\sigma(\bm{x})$, and $\langle\cdots\rangle $ denotes the average over the whole events.

\section{Kibble-Zurek scaling}
The cumulants of the sigma field are influenced by non-universal factors in the calculations, such as the relaxation time, the mapping between 3d Ising model and the hot QCD systems,  the evolution trajectory, etc.  Recently, it was realized that, within the framework of KZM, one could construct universal variables
that are independent from the non-universal factors~\cite{Mukherjee:2016kyu}. In this section, we will first explain
the Kibble-Zurek scaling near the critical point, and then explore it with the Langevin dynamics of model A.

\begin{figure}[tbp]
\center
\includegraphics[width=6.0 in]{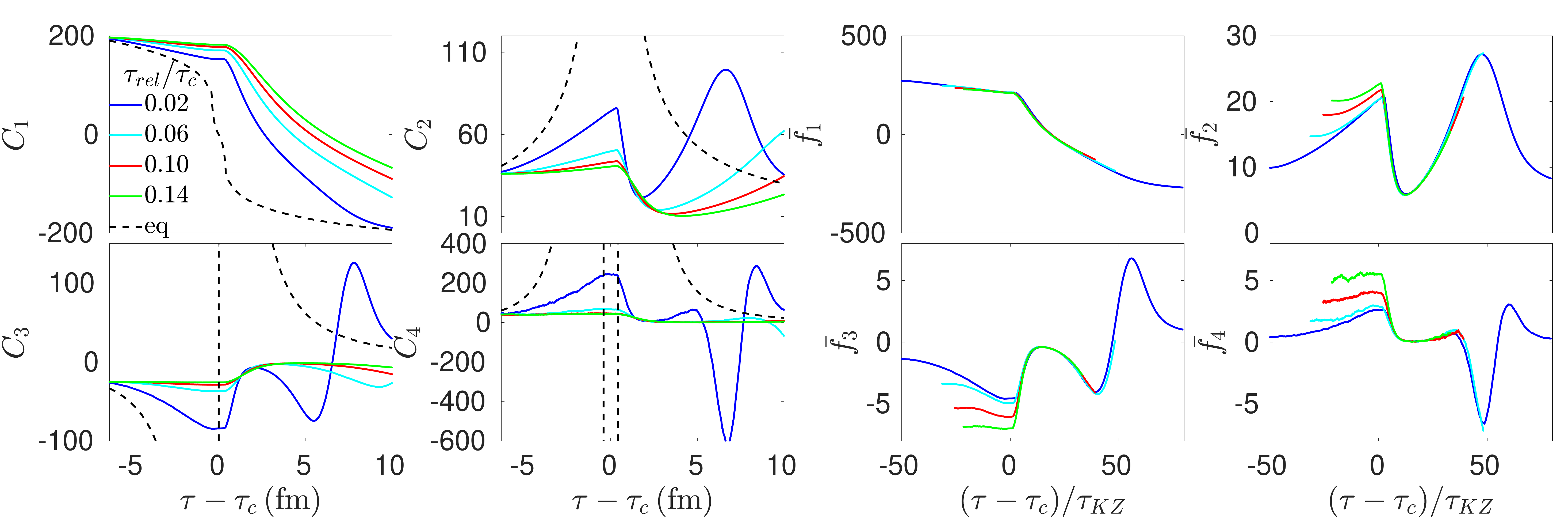}
\caption{(Color online) (Left): The evolution of the non-equilibrium cumulants $C_n$ (n=1 ... 4) along the trajectory of type A, with different relaxation time  $\tau_{rel}/\tau_c=0.02,0.06,0.1,0.14$. The black dashed line represents the equilibrium value. (Right): The constructed universal function $\bar{f}_n((\tau-\tau_c)/\tau_{\mbox{\tiny KZ}},\theta_{\mbox{\tiny KZ}})$ (n=1 ... 4) as a function of the rescaled time $(\tau-\tau_c)/\tau_{\mbox{\tiny KZ}}$. }
\label{TypeA}
\end{figure}
\begin{figure}[tbp]
\center
\includegraphics[width=6.0 in]{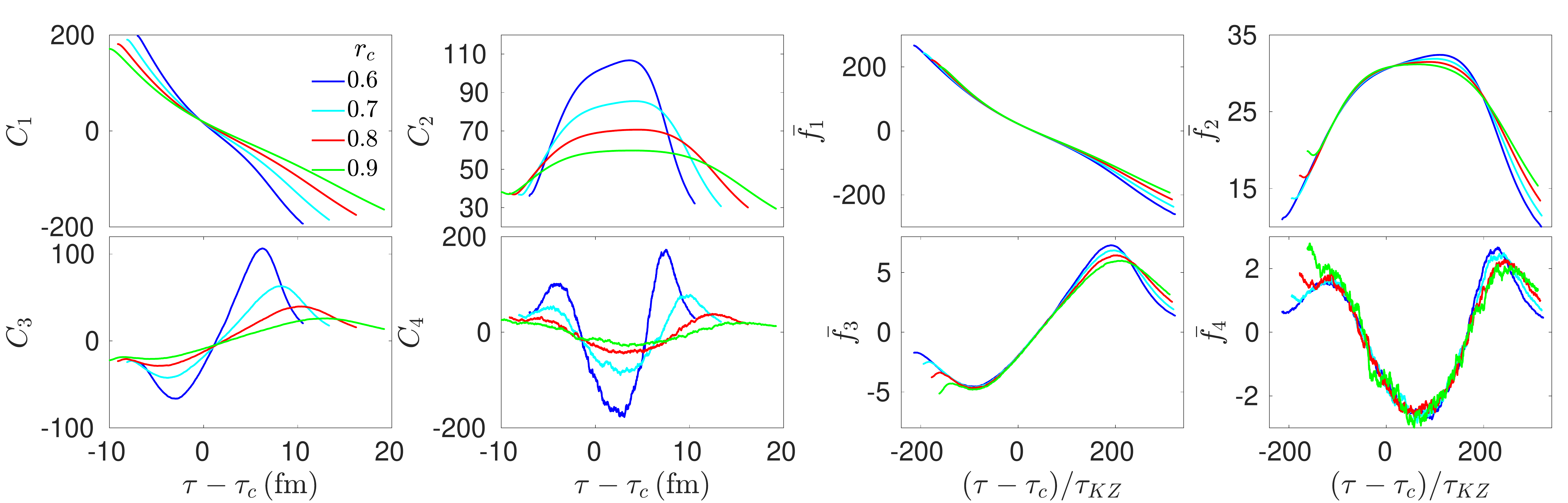}
\caption{(Color online) Similar to Fig.~2, but for the case of non-equilibrium evolutions along different trajectories of type B}
\label{TypeB}
\end{figure}

As the hot QCD system evolves near critical point, the effective relaxation time dramatically increases with the large enhancement of the correlation length: $\tau_{\mbox{\tiny eff}}= \tau_{\mbox{\tiny rel}}(\xi_{\mbox{\tiny eq}}/\xi_{\mbox{\tiny min}})^3$ . Meanwhile, the quench time $\tau_{\mbox{\tiny quench}}$ that measures the variation rate of the thermodynamic potential decreases with the fast expansion of the system. For the trajectory of type A, the quench time is mainly determined by the changing rate of the correlation length: $\tau^\xi_{\mbox{\tiny quench}}=|\xi_{\mbox{\tiny eq}}/(\partial_\tau \xi_{\mbox{\tiny eq}})|$. For the evolution along trajectories of type B, the changing rate of parameter $\theta$ is much larger than the one of the correlation length, which becomes the dominant factor to influence the variation of the potential. Correspondingly, the quench time is defined as: $\tau^\theta_{\mbox{\tiny quench}}=|\theta/\partial_\tau \theta|$. As illustrated in Fig.~\ref{fig2} (middle and right), there exists a specific point $\tau^*$ where $\tau_{\mbox{\tiny eff}}$ is equal to $\tau_{\mbox{\tiny quench}}$. As the system evolves to that point,  the order parameter field $\sigma$ is incapable of following the change of the thermodynamic potential due to the large relaxation time, which becomes "frozen". The relaxation time at $\tau^*$ that characterizes the typical emergent time scale is denoted as $\tau_{\mbox{\tiny eff}}(\tau^*)\equiv\tau_{\mbox{\tiny KZ}}$.  The correlation length $\xi_{eq}$ at $\tau^*$ that characterizes typical length scale of the correlated domain is denoted as $\xi_{\mbox{\tiny eq}}(\tau^*)\equiv l_{\mbox{\tiny KZ}}$. Similarly, the parameter $\theta$ at $\tau^*$ is denoted as: $\theta(\tau^*)\equiv\theta_{\mbox{\tiny KZ}}$. With $l_{\mbox{\tiny KZ}}$, $\tau_{\mbox{\tiny KZ}}$ and $\theta_{\mbox{\tiny KZ}}$, one can propose the following scaling form for the cumulants of the sigma field~\cite{Mukherjee:2016kyu}:
\begin{align}\label{scaling}
  C_n(\tau-\tau_c)) \sim l^{\frac{-1+5(n-1)}{2}}_{\mbox{\tiny KZ}}\bar{f}_n((\tau-\tau_c)/\tau_{\mbox{\tiny KZ}},\theta_{\mbox{\tiny KZ}}),\qquad n=1,2,3,4,\cdots.
\end{align}
where $\tau_c$ is the time when the hot QCD system evolves to the crossover phase transition line.  The exponent of $l_{\mbox{\tiny KZ}}$ is consistent with the $n-$order cumulants in equilibrium which is also proportional to the $[-1+5(n-1)]/2$ powers of the correlation length $\xi_{\mbox{\tiny eq}}$~\cite{Stephanov:2008qz}.
After rescaling  $C_n$ and $\tau-\tau_c$ with the Kibble-Zurek scale $l^{\frac{-1+5(n-1)}{2}}_{\mbox{\tiny KZ}}$ and  $\tau_{\mbox{\tiny KZ}}$, one could obtain the universal function $\bar{f}_n$.

The left panel of Fig.~\ref{TypeA} presents the non-equilibrium cumulants of the sigma field that evolve along the trajectory of type A
with differen relaxation time $\tau_{rel}/\tau_c=0.02, \ 0.06, \ 0.1$ and $0.14$. It shows that the relaxation time strongly influences the cumulants of the evolving sigma field. From Eq.~(\ref{scaling}), one could obtain $\bar{f}_n$ as a function of the rescaled time $(\tau-\tau_c)/\tau_{\mbox{\tiny KZ}}$ for each cumulant curve in Fig.~\ref{TypeA} (left). The right panel in Fig.~\ref{TypeA}  shows that, with such rescaling procedure, these different $C_n$  ($n=1...4$) curves converge into the universal curves $\bar{f}_n$  near the critical point. Note that, for trajectory of type A, the quenching time is determined by the changing rate of $\xi_{eq}$. Therefore, one can construct the universal function from the respective scales of $l_{\mbox{\tiny KZ}}$ and $\tau_{\mbox{\tiny KZ}}$.

Fig.~3 shows the time evolution of the non-equilibrium cumulants along these different trajectories of type B, which shows a strong dependence on the evolution trajectory. For these trajectories, the quenching time is determined by the changing rate of $\theta$. Therefore, the construction of the universal function $\bar{f}_n((\tau-\tau_c)/\tau_{\mbox{\tiny KZ}},\theta_{\mbox{\tiny KZ}})$ involves three different scales, $l_{\mbox{\tiny KZ}},\tau_{\mbox{\tiny KZ}}$ and $\theta_{\mbox{\tiny KZ}}$. Here, we tune the free parameter $\tau_{\mbox{\tiny rel}}/\tau_c$ to ensure that $\theta_{\mbox{\tiny KZ}}$  is the same for different trajectories. From Eq.(3.1), one could obtain the universal scaling function $\bar{f}_n$.  As shown in Fig.~\ref{TypeB}, after such rescaling procedure, these different cumulants curves becomes
universal ones in the vicinity of the crossover line, which are independent from the evolving trajectories.

\section{Summary}

The non-equilibrium critical fluctuations of the hot QCD system in the model calculations are always influenced by many non-universal factors, such as the relaxation time, the mapping between 3d Ising model and the hot QCD system, the evolution trajectory, etc. Within the framework of Kibble-Zurek Mechanism, we studied the universal behavior of the critical fluctuations using the Langevin dynamics of model A.  We mainly focused on two cases: 1) systems evolving along one trajectory with fixed chemical potential (type A), but with different relaxation times, 2) systems evolving along different trajectories with different approximate-equal correlation length near the phase transition (type B).
After locating
the typical scales of time, length and angle  $\tau_{\mbox{\tiny KZ}}$, $l_{\mbox{\tiny KZ}}$, and $\theta_{\mbox{\tiny KZ}}$, we constructed the related universal scaling function $\bar{f}_n((\tau-\tau_c)/\tau_{\mbox{\tiny KZ}},\theta_{\mbox{\tiny KZ}})$ for the cumulants of the sigma field. The related numerical simulations showed that
such universal functions have been successfully constructed for both cases, which are either independent on the relaxation times or the evolution trajectories. In the near future, it is worthwhile to explore the construction of possible universal observables near the QCD critical point for the RHIC BES program and to investigate whether the realistic QGP fireball evolution and particle emissions could preserve such universal behavior.

\section{Acknowledgments}

The authors acknowledge the fruitful discussion with S. Mukherjee, M. Stephanov and Y. Yin,
This work is supported by the NSFC and the MOST under grant
Nos.11435001, 11675004 and 2015CB856900.

\end{document}